# Neuromorphic Imaging Flow Cytometry combined with Adaptive Recurrent Spiking Neural Networks

Georgios Moustakas,[1] Ioannis Tsilikas,[2] Adonis Bogris[1] and Charis Mesaritakis[3,*]

[1]University of West Attica, Dept. of Informatics and Computer Engineering, Aghiou Spiridonos, 12243, Egaleo, Athens, Greece.
[2]Department of Physics, School of Applied Mathematical and Physical Sciences, Zografou Campus, 157 80 Athens, Greece.
[3]Department of Biomedical Engineering, University of West Attica, Aghiou Spiridonos, 12243, Egaleo, Athens, Greece
*cmesar@uniwa.gr

**Abstract:** We present an experimental imaging flow cytometer using a 1 μs temporal resolution event-based CMOS camera, with data processed by adaptive feedforward and recurrent spiking neural networks. Our study classifies PMMA particles (12, 16, 20 μm) flowing at 0.7 m/s in a microfluidic channel. Processing of experimental data highlighted that spiking recurrent networks, including LSTM and GRU models, achieved 98.4% accuracy by leveraging temporal dependencies. Additionally, adaptation mechanisms in lightweight feedforward spiking networks improved accuracy by 4.3%. This work outlines a technological roadmap for neuromorphic-assisted biomedical applications, enhancing classification performance while maintaining low latency and sparsity.

## 1. Introduction

Imaging flow cytometry (IFC) is a cellular analysis technique that combines the features of flow cytometry and confocal microscopy to provide rich morphological information, such as cell size and shape, alongside fluorescence data at single-cell resolution or across a population of cells [1,2]. The key advantage of IFC over standard microscopy is the ability to record moving cells without special sample preparation that can hinder cell morphology and increase the lab-to-result latency. Due to this key feature, IFC has infiltrated multiple biomedical applications ranging from detection of cell phenotypes [2], drug discovery [3] or DNA sequencing [4].

On the other hand, IFC schemes are plagued by two issues. The first is the limited spatial resolution, due to motion blur [5], when high particle speed is utilized. This impediment stems from the limited exposure time at the typical CMOS cameras used for recording. A second issue that emerges from the need for high-speed operation, is the high data volume in IFCs that puts stringent requirements to both the data storing and processing back-end.

Aiming to address the first issue, multiple approaches have emerged such as time-delayed and integrate cameras that can provide high resolution and multiplexed imaging at operation speeds up to 3000 particles/sec [5]. TDI cameras have also been used in conjuction with spectral flow cytometry for the immunophenotyping of low-abundance cells [6]. In addition, streak imaging have been used for detection of rare cells with flow rates up to 10 mL/min [7]. Unfortunately, the majority of these solutions are bulky or/and expensive and thus do not allow for miniaturized, low-cost integrated systems. Regarding the second impediment; namely large data volume analysis, one potential solution consists of using an expert who can accurately interpret the extracted morphological features manually [8] which is a less practical approach when the data volume increases at a high level. More recently a more robust solution comprises the use of sophisticated deep machine learning models. The latter approach imposes its own restrictions; namely, it requires large neural models and high-end digital platforms for neural inference and most importantly, demands extensive training. In this context, the sophisticated digital back-end is also hard to integrate.

An unconventional solution to both these impediments are event cameras, which are biologically inspired asynchronous sensors that measure changes in brightness of each pixel independently, offering a high temporal resolution, high dynamic range, low latency, ultra-low power consumption and more critically, data represented in a sparse stream of events rather than a redundant grid of pixel values [9]. More specifically, whenever a change in brightness exceeds a predefined threshold, an event $E(X,Y,T,P)$ is generated. This event contains the location of the pixel ($X$, $Y$ coordinates), the timestamp ($T$), and a 1-bit value indicating the polarity, which denotes whether the brightness has increased or decreased. Since only changes in brightness are transmitted instead of entire frames, redundant data are minimized, background information is intrinsically removed, and power is conserved by processing only pixels associated with events.

Event-based sensors have found diverse applications in fields such as 3D reconstruction, simultaneous localization and mapping and Optical Flow Estimation [9]. Obviously IFC is an application that has already exploited the merits of event-based sensors. IFC with event-based cameras and machine learning algorithms was realized in [10] where fungal cells were classified and clustered utilizing unsupervised learning paradigm offering up to 100% accuracy on the classification of fungal cells, without relying on training labels. High flow IFC setup has also been combined with lightweight feedforward network (FNN) and a more complex recurrent neural network (RNN) to classify artificial particles [11]. In this case, the spike-generating nature of the sensor was not fully exploited and the back-end machine learning models dictated the use of synthetic frames (images).

More recently, in [12] an IFC scheme was combined with a photonic neuromorphic accelerator prior to typical light-weight machine learning models, allowing the reduction of trainable parameters, but again typical image-based data representation was employed. In [13], event-based IFC has been firstly combined with an offline software based spiking neural network (SNN torch), aiming to exploit the asynchronous nature of the incoming data and reduce the overall processing latency and power consumption. In this case, a sophisticated data pre-processing was used so as to extract features from the spiking data, accompanied by a neuron-hungry SNN scheme. Furthermore, in [14], the event-based camera is merged with a spiking neural network, implemented on a dedicated neuromorphic hardware. The same dataset with [13] was utilized. This approach achieved enhanced accuracy, surpassing a frame-based model trained on the same dataset. Lastly, in [15], the task of IFC cell classification and localization was addressed by developing a variant of YOLO v3 with a modified backbone, consisting of convolutional and batch normalization layers alongside Leaky-Integrate-Fire (LIF) neurons. This model outperformed the traditional YOLO v3 model at the cost of a cumbersome model with a significant number of trainable parameters.

Although, there are several works in the literature showing the potential of spiking processing in event-based systems [13] the potential of lightweight SNN models in terms of the classification and complexity performance is largely unexplored. Hence the detailed comparison of different SNN models that could be implemented in dedicated neuromorphic hardware offering real-time detection and classification is still a pending issue. Towards this direction, in this work, we estimate the classification accuracy and compare the computational complexity of various lightweight SNN schemes with and without the adaptation mechanism, trained on experimental data generated by an event-based IFC system similar to [13,14].

In particular, we have considered Spiking Multilayer Perceptron (SMLP) based on leaky integrate and fire (LIF) neurons with and without adaptation mechanism, Spiking Long Short Term Memory (SLSTM) and Spiking Gated Recurrent Unit (SGRU) and compared their classification accuracy and computational complexity with respect to the number of real multiplications. We have found that the majority of the networks can surpass a classification accuracy 92% at all scenarios with SLSTM reaching the highest accuracy of 98.43% followed by SGRU. Interestingly, in this work we investigate performance of a second order adaptive LIF, compared to typical SMLPs. The adaptation mechanism, without affecting complexity raises accuracy to 98.18 and 96.8% respectively. In terms of complexity, the number of real multiplications varies from 5,035,600 to 60,600 for the SLSTM and first-order SMLP. However, it should be taken into consideration that, due to the binary nature of the input data, multiplications can be reduced to accumulations, thereby decreasing the complexity. The results show that the adaptation mechanism is able to increase classification accuracy across all three data insertion scenarios. On the other hand, recurrent networks can encode temporal features more effectively than any other variant, achieving the highest classification accuracy in every scenario, but at the expense of complexity increase.

## 2. Experimental Setup and Numerical Tools

### 2.1. Experimental Setup

The experimental setup is depicted in Fig. 1. It is a low-cost light-emitting diode (LED)-based IFC scheme utilizing an event-based camera (Prophesee Metavision EVK4 [16]) and a generic micro-fluidic setup. More specifically, two microscope objectives were installed to focus and gather light into/from the microfluidic channel while the 12 μm, 16 μm, and 20 μm Polymethyl Methacrylate (PMMA) particles are delivered into a microfluidic chip featuring a single 100 x 100 μm channel. This process is controlled by a vacuum pump, which maintains a steady flow and enables the particles to achieve mean velocities ranging from 1 to 0.7 m/s, analogous to a particle flow rate of 500 to 350 particles/s. The event-based camera supports a spatial resolution of 640x480 pixels and a temporal resolution of 1 μs.

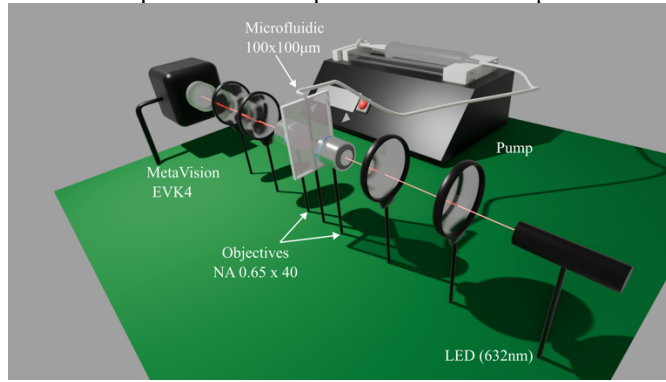

Figure 1. Proposed IFC, illustration from [12]

### 2.2. Data Preprocessing

Event-based cameras emit tuples $E(X,Y,T,P)$ that include the coordinates $(X,Y)$, a timestamp and the polarity that corresponds to the value +1 or -1 depending on whether pixels' intensity is increasing or decreasing respectively. Event count $I(X,Y)$ on each coordinate is then calculated by summing all the events per pixel (without considering the polarity) over an integration window T = 3 msec , which is set according to the particle's speed [12].

The event count $I(X,Y)$ can be viewed as a form of rate coding, as higher event counts indicate increased spiking activity at a specific spatial coordinate within the integration time. Since the microfluidic channel is generic rather than custom, it does not restrain the position of the particles and is unable to align them along a specific trajectory. For this reason, a lightweight center of mass tracking algorithm was employed. Following this step, the frame was cropped to a 100x100 pixel region that contained particle relevant events. Finally, each synthetic frame was normalized using the detected width of the microfluidic channel during each experimental session so as to compensate for out of focus measurements.

The resulting data frames, despite being in a rate coding format, cannot be fed directly into a spiking neural network since each *(X,Y)* coordinate contains the event count and not a discrete binary value X ∈ {0,1}. To mitigate this, *I(X,Y)* was clipped to ensure that its contents lie within the range [0,1], yielding a binary encoding of the data.

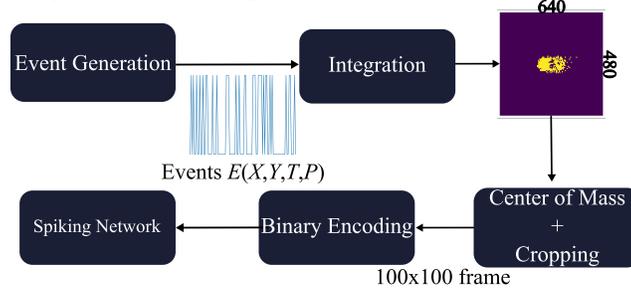

Figure 2. Data preprocessing pipeline

Based on this processing pipeline, an unbalanced dataset of 4389 unique frames was prepared out of which 1216 particles belonged to 20 μm class, 1811 to 16 μm and 1351 to 12 μm respectively. To minimize experimental bias, data were collected across multiple sessions using different microfluidic chips.

### *2.3. Spiking Models*

Spiking Neural Networks (SNNs) are biologically inspired computational models that have recently started been used alongside event-based sensors [9]. The rationale of this merge is that SNNs can leverage the sparsity of event-based data while typically offering lower power consumption when executed on specialized neuromorphic hardware, compared to conventional digital counterparts [17]. Unlike Artificial Neural Networks (ANNs), which rely on sigmoid or other activation functions [18], Spiking Neural Networks (SNNs) integrate their weighted inputs to regulate their internal state, known as the membrane potential. When the membrane potential exceeds a threshold, the neuron becomes excited and generates an action potential, which propagates through its axon to subsequent neurons. In addition, this spike-based approach replaces high-precision multiplications between layer activations and weights with multiplication between spikes and weights, thereby reducing the overall complexity.

In this work, we trained various spiking architectures, ranging from SMLP to recurrent models, including SLSTM and SGRU networks. Adaptation mechanism [6] was implemented in the second-order SMLP architecture. However, the basic building blocks of all these networks are first and second order leaky integrate and fire neurons. Hence, before shedding light to the implementation details, a brief introduction to these models along with adaptation mechanism used in this context is essential.

### 2.3.1 Leaky Integrate and Fire Model

The most straightforward neuronal model is the leaky integrate and fire model whose dynamics can be modeled with Resistor-Capacitor (RC) circuit. The membrane potential is formulated as

$$\tau_{mem} \frac{dU(t)}{dt} = -U(t) + I(t) \cdot R \quad (1)$$

Where *I(t)* is the input current following the terminology in [17] (input data), *R* the resistance and $\tau_{mem}$ is the RC time constant. It is immediately evident from (1) that membrane potential acts as an integrator of the input current. Neuron becomes excited and communicates spikes to other neurons when membrane potential exceeds a threshold θ. After an action potential, membrane potential is reset by subtracting the value θ from $U(t)$, something not included in (1). To be able to numerically solve this first order ordinary differential equation (ODE), forward Euler method is used with $\Delta t = 1 \ll \tau_{mem}$ to yield [17]:

$$U[t+1] = \beta U[t] + (1-\beta)I[t+1] \quad (2)$$

where $\beta = e^{-\frac{\Delta t}{\tau_{mem}}}$. By incorporating $(1-\beta)$ into a learnable matrix W and including the reset term, eq. (2) can be reformulated as [17]:

$$U[t+1] = \beta U[t] + W \cdot I[t+1] - S[t]\theta \quad (3)$$

where S[t] is given by (4):

$$S[t] = \begin{cases} 1, & U[t] > \theta \\ 0, & \text{otherwise} \end{cases} \quad (4)$$

Hence, unlike other neuronal models which directly reset the membrane potential to a value $U_{rest}$, reset by subtraction in this model is realized by (3). A typical output of such a model is shown in Fig. 3a. It is shown that each input value has a direct contribution to the membrane potential which resets whenever it exceeds the threshold and decays back to zero in the absence of any input

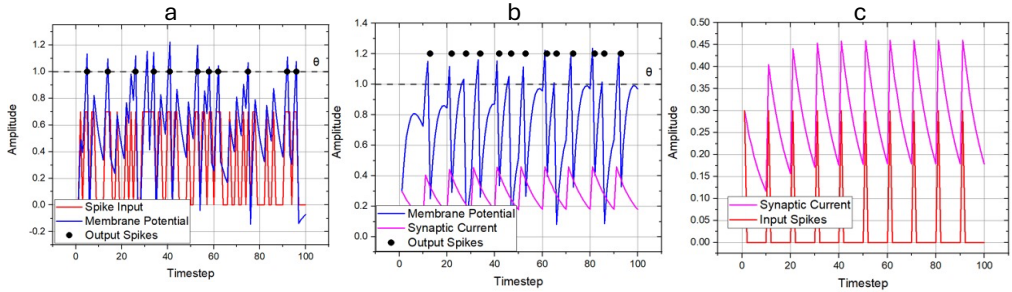

Figure 3. (a) Typical output of LIF neuron with β = 0.8 and θ = 1 (b) Membrane Potential & Synaptic Current of second order LIF modeled with β = 0.8 and α = 0.9 (c) Synaptic current and input of second order LIF modeled using β = 0.8 and α = 0.9

### 2.3.2 Second order Leaky Integrate and Fire Model

Previous models assumed that input current has immediate/direct influence on the membrane potential. A more realistic model is the Synaptic [17] or Second order Leaky Integrate and fire model which accounts for the gradual temporal dynamics of input current. In this model, the state of the neuron incorporates another decaying term, $I_{syn}(t)$ used to model the synaptic input with a decay rate equal to $a = e^{-\frac{\Delta t}{\tau_{syn}}}$. The dynamics of the model are governed by [17]:

$$I_{syn}[t+1] = \alpha I_{syn}(t) + W \cdot X[t+1] \quad (5)$$
$$U[t+1] = \beta U[t] + I_{syn}[t+1] - S[t]\theta \quad (6)$$

As shown in Fig. 3b and Fig. 3c, synaptic current $(I_{syn}(t))$ integrates the input current and decays at rate $a$ while membrane potential instead of integrating the input current, integrates the synaptic current and decays as before, with a decay rate β.

### 2.3.3 Adaptation Mechanism

While higher-order models, such as the Hodgkin-Huxley model combined with the muscarinic potassium channel [6], include adaptation, a simpler approach was adopted in this work to model adaptive threshold. In this method, an additional ODE is introduced to control the threshold, which is no longer a constant value but instead a dynamic variable as:

$$\theta[t+1] = \gamma\theta[t] + S[t] \quad (7)$$

As in the other models, the decay rate of threshold is given by $\gamma = e^{-\frac{\Delta t}{\tau_{thr}}}$. For simplicity, the dynamics of Adaptive LIF are shown in Figures 4a and 4b where it is demonstrated that the threshold dynamically increases every time there is an output spike and decays when there is no output spike. Membrane potential behaves as in the typical LIF model.

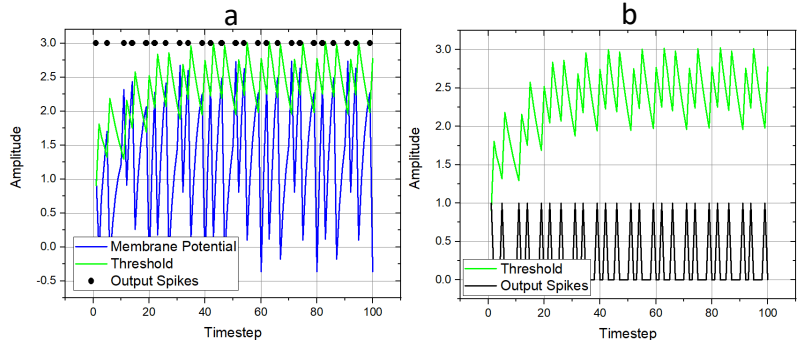

Figure 4 (a) Membrane Potential and Threshold of second order Adaptive LIF modeled using β = 0.8 and α = 0.9 and γ = 0.9 (b) Threshold and output spikes of second order Adaptive LIF modeled using β = 0.8 and α = 0.9 and γ = 0.9

### 2.3.4 Spiking Long Short Term Memory (SLSTM)

Spiking Long Short Term Memory model [19] is the spiking version of LSTM [20] network with spike activations in the hidden state of the model that is used as membrane potential while the cell state is used as a synaptic current. This model does not have any decay mechanism since the LSTM cell state acts as a long short-term memory and is used to regulate the flow of information in the cell, just like in conventional LSTM networks. Hence, SLSTM can be formulated as [17,19]:

$$i[t] = \sigma(W_{ix}x[t] + b_{ix} + W_{ih}U[t-1] + b_{ih})$$
$$f[t] = \sigma(W_{fx}x[t] + b_{fx} + W_{fh}U[t-1] + b_{fh})$$
$$g[t] = \sigma(W_{gx}x[t] + b_{gx} + W_{gh}U[t-1] + b_{gh})$$
$$o[t] = \sigma(W_{ox}x[t] + b_{ox} + W_{oh}U[t-1] + b_{oh})$$
$$I[t] = f[t] \odot I[t-1] + i[t] \odot g[t] \quad (8)$$
$$U[t] = o[t] \odot tanh(I[t]) - S[t]\theta \quad (9)$$

Where W are the weight matrices for the gating units, ⊙ is the Hadamard product, b are bias vectors, i[t], o[t] are gating units and g[t] is candidate memory of LSTM, or $\tilde{c}$.

### 2.3.5 Spiking Gated Recurrent Unit (SGRU)

In the same manner as SLSTM, spiking GRU is another spiking recurrent model built on top of Gated Recurrent Unit (GRU) [21] network defined as :

$$z[t] = \sigma(W_z x[t] + G_z h[t-1])$$
$$r[t] = \sigma(W_r x[t] + G_r h[t-1])$$
$$\tilde{h}[t] = tanh(W_h x[t] + r[t] \odot G_h h[t-1])$$
$$h[t] = z[t] \odot h[t-1] + (1 - z[t]) \odot \tilde{h}[t]$$
$$U[t] = h[t] - S[t]\theta \quad (10)$$

Where σ is the sigmoid activation function, $W, G$ are the weight matrices (without bias vectors) and z and r denote the update and reset gates respectively. Output and intermediate memory are denoted by h and $\tilde{h}$. Like SLSTM, the output of the GRU is used as membrane potential. While the synaptic current is not modeled in this case, as the GRU is internally able to remove redundant information using internal gates.

*2.4. Spiking Network Architectures*

The neuronal models introduced in the previous section, except SLSTM and SGRU, cannot be used directly in their current form because, within the context of deep learning, they function merely as nonlinear activation functions. To enable training, they must be integrated into more complex architectures.

2.4.1 Spiking Multilayer Perceptron (SMLP)

Spiking multilayer perceptron consists of a linear layer used to model the synaptic connections followed by three spiking neurons. These neurons can be either first-order or second-order and may exhibit adaptive or non-adaptive behavior, the network is shown in Fig 5a. The inputs are multiplied by a synaptic weight matrix and then passed directly to a classifier which is going to perform the classification on the particles.

The number of trainable parameters depends mostly on the size of the synaptic weight matrix (linear layer) and can be calculated by:

$$params_{MLP} = x_i \cdot n_o + n_o \quad (11)$$

Where $x_i$ are the input features and $n_o$ are the output neurons. Based on this topology, three models with the same number of synaptic weight parameters were utilized (table I), including first-order LIF, second-order LIF with adaptation mechanism and second-order LIF without adaptation. The adaptive second-order and non-adaptive second-order models have the same number of parameters. This is because the former includes three decay rates as trainable parameters, while the latter has two decay rates and the threshold as trainable parameters.

**TABLE I**

**Number of parameters**

| Network | Rows/Columns* | Rows & Columns* |
| --- | --- | --- |
| LIF SMLP | 305 | 603 |
| Synaptic SMLP | 306 | 606 |
| Adaptive Synaptic SMLP | 306 | 606 |
| SRGU + SMLP | 22.956 | 37.956 |
| SLSTM + SMLP | 30.556 | 50.556 |

* The integration is done with respect to rows, columns or both.

2.4.1 Spiking Recurrent + MLP

The previous architecture, though lightweight, fails to capture the temporal relations of the features due to lack of any model memory. The LIF layer does not account for information from adjacent neurons; instead, each neuron is isolated and relies solely on its own history of signals. Recurrent architectures, on the other hand, leverage their internal state to efficiently capture temporal dependencies, making them more efficient than conventional ANNs in this regard. Hence, adding a spiking recurrent unit before the MLP could enrich the classifier with temporal features, thereby enhancing the model's accuracy. Network is depicted in Fig 5b, where 50 neurons are used in all the recurrent architectures.

The number of parameters is not as straightforward as in the MLP case since the network contains a recurrent SNN that can vary depending on the implementation. In this work, we mainly employ SGRU and SLSTM as the recurrent layer which in turn contains GRU and LSTM cells.

**SGRU + MLP**: As explained in the previous section, SGRU employs a GRU cell. Hence the number of trainable parameters is given as:

$$params_{SGRU+MLP} = params_{GRU} + params_{MLP}$$
$$= (3 \cdot n_h \cdot n_i + 3 \cdot n_h \cdot n_h + 6 \cdot n_h) + n_h \cdot n_o + n_o \quad (12)$$

**LSTM + MLP**: Equivalently, the number of parameters for this network is formulated as:

$$params_{SLSTM+MLP} = params_{LSTM} + params_{MLP}$$
$$= (4 \cdot n_h \cdot n_i + 4 \cdot n_h \cdot n_h + 8 \cdot n_h) + n_h \cdot n_o + n_o \quad (13)$$

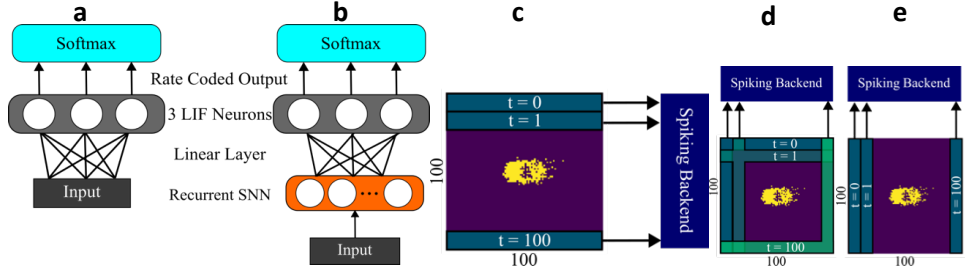

Figure 5 (a) SMLP Architecture, (b) Recurrent SMLP architecture (c) Row as input method (d) Rows and columns as inputs (e) column as input

### 2.4. Data Insertion Process

Before discussing the complexity of each topology, it is essential to distinguish between the three data insertion strategies employed, as they directly influence the number of real multiplications. In all three methods, we iterate over either rows, columns or both, so the number of timesteps used in the integration is t=100.

The first method is illustrated in Fig. 5c, where each row is used as input to the spiking network, enabling the integration of columns. As the particle moves from right to left, this method captures less temporal information between timesteps. The second method, shown in Fig. 5e, involves inserting each column separately into the spiking network, allowing for the integration of rows. As demonstrated in the next section, this method captures the temporal dependencies in the vertical direction, which is linked to particle size and is not affected by variation in the particle speed (the axis of motion coincides with the rows). Finally, the third method concatenates each row and column before feeding the combined input into the neural network, enabling a diagonal scan across the particle frame, as shown in Fig 5d. This approach extracts a greater number of features and encodes the temporal relationships between timesteps in a more efficient manner.

## 3. Training & Complexity Analysis

### 3.1. Model Training

All models were trained and evaluated using the PyTorch [22] GPU backend on Nvidia Titan RTX GPU with 24 GB of VRAM. Synaptic weights were modeled using Linear layers provided by PyTorch, along with GRU and LSTM cells. Simulations of spiking neuronal models were conducted using SnnTorch [17]. The loss function used during optimization was the Cross Entropy Spike Rate, as defined in [17]. Weight optimization was carried out using the Adam optimizer with the same parameters as specified in [23]. The dataset was divided into 70% for training and 30% for testing, with an empirically determined batch size of 64. The number of optimization iterations (epochs) was set to 150. To overcome the non-differentiability of the Heaviside function, it was retained as-is during the forward pass but during the backward pass, its derivative was replaced with the gradient of the shifted arc-tan function [24]:

$$S = \frac{1}{\pi} arctan\left(\pi U \frac{\alpha}{2}\right) \tag{14}$$

$$\frac{dS}{dU} = \frac{1}{\pi} \frac{1}{1 + \left(\pi U \frac{a}{2}\right)^2} \tag{15}$$

In all models, the decay rate was set as trainable parameter while in non-adaptive models, the threshold was also set as a trainable parameter.

**TABLE II**

**Accuracy and number of multiplications of each network**

|  | Accuracy in % | | | Number of Multiplications | |
| --- | --- | --- | --- | --- | --- |
| Network | Rows* | Columns* | Rows & Columns* | Rows or Columns* | Rows & Columns* |
| LIF SMLP | 90.88 | 88.45 | 96.354 | 30.600 | 60.600 |
| Synaptic SMLP | 90.6 | 92.1 | 96.4 | 30.900 | 60.900 |
| Adaptive Synaptic SMLP | 93.22 | 92.787 | 96.875 | 31.200 | 61.200 |
| SRGU + SMLP | 94.31 | 93.48 | 98.18 | 2.285.600 | 3.785.600 |
| SLSTM + SMLP | 96.61 | 95.57 | 98.43 | 3.035.600 | 5.035.600 |

* The integration is done with respect to rows, columns or both

### 3.2. Complexity Analysis

When designing such networks for embedded devices or FPGAs rather than specialized neuromorphic processors, it is crucial to evaluate the computational complexity in terms of real multiplications (RM). However it should be noted that since the input is entirely binary all multiplications can be reduced to accumulations, effectively reducing the hardware resources and power consumption [17,25]. The following process focuses entirely on the required multiplications per sample in the inference stage. Addition is excluded because it is regarded a computationally inexpensive operation [26]. The number of real multiplications is summarized in Table II for each network topology and data insertion method.

**Linear + Leaky (Leaky SMLP):** Leaky SMLP is the leaky version of the SMLP and comprises a linear layer followed by a LIF. Eq. (3) includes 3 major multiplication operations, two of them are elementwise and one of them is matrix multiplication. The elementwise multiplications $\beta U[t]$, $S[t]\theta$ require approximately $n_o$ multiplications while $W \cdot I[t+1]$ is performed by the Linear layer and requires $n_i \cdot n_o$ multiplications where $n_i$ are the input features and $n_o$ is the number of neurons. The total number of RMs is given by:

$$RM_{LSMLP} = n_s \cdot (n_i \cdot n_o + 2 \cdot n_o) \tag{16}$$

Where $n_s$ is the number of timesteps used.

**Linear + Adaptive Synaptic (AdaSynaptic SMLP):** Adaptive Synaptic SMLP is the synaptic (second order) version of the SMLP with adaptation mechanism. Using equations (5), (6) and (7) we now see 4 elementwise multiplications $\beta U[t], S[t]\theta[t], \alpha I_{syn}(t), \gamma\theta(t)$, and 1 matrix multiplication which again is computed using the Linear layer. Hence the number of real multiplications is

$$RM_{AdaSynSMLP} = n_s \cdot (n_i \cdot n_o + 4 \cdot n_o) \tag{17}$$

**Linear + Synaptic LIF (Synaptic SMLP):** Synaptic SMLP is the second order leaky integrate and fire version of the SMLP. Using (5) and (6), it is clear that one more elementwise multiplication is added to the Linear + Leaky case due to the existence of the synaptic current mechanism. The total number of multiplications is:

$$RM_{SynapticMLP} = n_s \cdot (n_i \cdot n_o + 3 \cdot n_o) \qquad (18)$$

**SGRU + MLP:** Spiking Gated Recurrent unit and MLP is the GRU based recurrent architecture mentioned above. GRU cell consists of 3 gating units where each requires $n_h \cdot n_i + n_h \cdot n_h$ multiplications. Adding the Hadamard products yields another $n_h$ multiplications, hence for all 3 gates, the total number of multiplications is

$$RM_{GRU} = n_h \cdot (3 \cdot n_i + 3n_h + 3) \qquad (19)$$

Spiking GRU adds one more elementwise multiplication due to the reset term ($S[t] \cdot \theta$) so the number of real multiplications required by the SGRU are

$$RM_{SGRU} = n_h \cdot (3 \cdot n_i + 3n_h + 3) + n_h \qquad (20)$$

Adding the SMLP classifier yields another $n_h \cdot n_o + 2 \cdot n_o$ multiplications, so the total RM of SGRU + MLP are:

$$RM_{SGRU+MLP} =$$
$$n_s \cdot (n_h \cdot (3 \cdot n_i + 3n_h + 3) + n_h + n_h \cdot n_o + 2 \cdot n_o) \qquad (21)$$

**SLSTM + MLP:** Spiking Long Short-Term Memory and MLP is the LSTM based architecture. LSTM cell includes 4 gating units where each requires $n_h \cdot n_i + n_h \cdot n_h$. Adding the 3 Hadamard products of $n_h$ results in a total number of RM:

$$RM_{LSTM} = n_h \cdot (4 \cdot n_i + 4 \cdot n_h + 3) \qquad (22)$$

SLSTM includes one more term due to the reset, same as the SGRU network, hence the total number of RM for SLSTM are

$$RM_{SLSTM} = n_h \cdot (4 \cdot n_i + 4 \cdot n_h + 3) + n_h \qquad (23)$$

Accounting for the SMLP classifier, SLSTM with MLP yield:

$$RM_{SLSTM+MLP} =$$
$$n_s \cdot (n_h \cdot (4 \cdot n_i + 4n_h + 3) + n_h + n_h \cdot n_o + 2 \cdot n_o) \qquad (24)$$

## 4. Results and Discussion

### 4.1. Results

Table II summarizes the accuracy, expressed as a percentage (%), for each network trained with all three integration methods: rows, columns, and rows & columns. The Spiking LSTM network achieves the highest accuracy of 98.43% when integrating both rows and columns, followed by the Spiking GRU, which achieves an accuracy of 98.18% with the same integration method. Among the SMLP variants, the adaptive second order LIF variant achieves the highest accuracy of 96.875%.

For other integration techniques, the Spiking LSTM achieves the highest accuracy of 95.57% when integrating columns, with the Spiking GRU achieving the second-highest accuracy of 93.48%. Among the SMLP variants, the Second-Order Adaptive LIF achieved the highest accuracy of 92.787%, followed by the other variants with accuracies of 92.1% and 88.45%, respectively

Finally, when integrating the rows of the particle, the Spiking LSTM achieves the highest accuracy of 96.61%, followed by the Spiking GRU which achieves an accuracy of 94.31%. Again, among the SMLP variants, the Second-Order Adaptive LIF outperforms the other two, achieving an accuracy of 93.22%, compared to 90.88% and 90.6%, achieved by LIF and Synaptic LIF SMLP respectively. A confusion matrix was created for the Spiking LSTM variant when integrating both rows and columns based on network predictions and ground truth labels as shown in Fig. 6a. All classes show a high percentage of accuracy (> 98%), reaching 1 for the smallest class of samples (12 μm). Finally, the extracted features of the LSTM variant were further processed using the Uniform Manifold Approximation and Projection [27] (UMAP) algorithm, as in [15], to evaluate the feature extraction ability and the correlation between the three classes. The results in Fig. 6b demonstrate the clusters formed by each particle class, with a few outliers from class 20 um appearing in the cluster for class 12 um. Models with more parameters and multiplications can reach higher accuracy than the simpler ones coming at a cost of both memory and computational complexity. For instance, the LSTM variant when integrating both rows and columns has 50.556 trainable parameters and executes 5.035.600 RM while the LIF SMLP has only 305 trainable variables and executes 60.600 RM with a degradation of around 2% in the accuracy. The Spiking GRU exhibits an accuracy in between LIF SMLP and SLSTM with 37.956 parameters and 3.785.600 RM while the other SMLP variants have the same memory and computational complexity as LIF SMLP. However, due to the binary nature of the input containing 1 spike per event, energy consumption can be reduced by a factor of 1.72 [28].

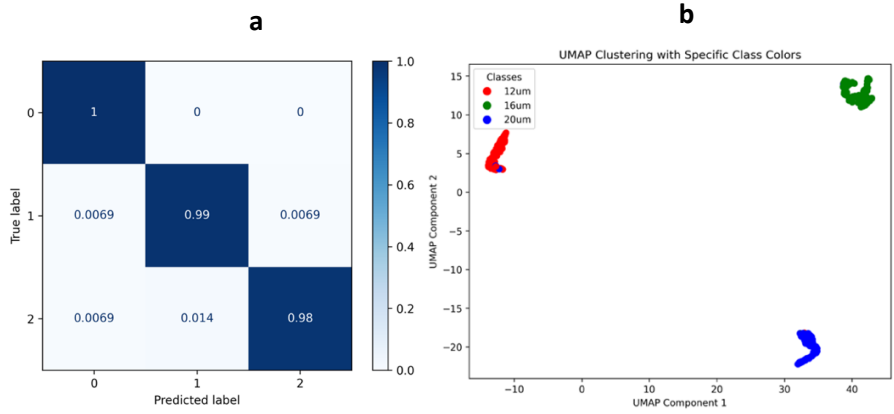

Figure 6 (a) Clusters formed in UMAP (b) SLSTM Confusion Matrix

## *4.2. Discussion*

The investigated models were able to achieve high levels of accuracy ( > 95%) when using both rows and columns as inputs. In all variants, this is due to the fact that this provides maximal temporal information to the model compared to the other two methods which isolate only a certain part -either row or column- on each timestep. More specifically, GRU and LSTM recurrent models, through the use of gating units, are better able to encode and retain relevant information during training compared to the short-term memory provided by the Leaky LIF, whose state is influenced only by the previous timestep, similar to a vanilla RNN architecture. The combination of synaptic current and threshold adaptation mechanism does not appear to significantly improve accuracy. This may be because second-order models offer no notable advantage over first-order models when temporal relationships occur over short time scales [3]. This is evident in the relatively small size of the particles in each class, especially when compared to the bounding box size of 100x100 pixels, where contiguous events do not persist over many timesteps.

In the other two data insertion methods, LSTMs and GRUs once again outperform the other models, owing to their gating units. The Second-Order Adaptive LIF variant achieves approximately 3% higher accuracy than the First-Order LIF model. This improvement is attributed to the inclusion of synaptic current and adaptive threshold mechanisms, which enhance the model's memory by allowing it to retain information across more timesteps. In contrast, the First-Order LIF model, as discussed in the previous paragraph, can only pass its state to the next timestep.

The difference observed across all models between the row-wise and column-wise methods could be attributed to the way temporal information is encoded. Since the particles flow from right to left, feeding the data column-wise captures more temporal information. Conversely, a row-wise approach would be more effective if the particles flowed from top to bottom.

Overall, the recurrent variants achieve higher accuracy compared to MLP architectures. However, this improvement comes at the expense of a significant increase in both the number of parameters and the number of real multiplications due to their implementation, which is based on PyTorch's LSTM and GRU cells. An alternative approach to utilizing recurrence could involve feeding the output of each layer back into its input or implementing a more sophisticated threshold adaptation mechanism—an area not explored in this work.

## 5. Conclusion

Event-based sensors, owing to their high temporal resolution, high dynamic range, and low power consumption, have found applications across a wide range of fields [9]. Imaging flow cytometry is another field that could benefit from this event-based sensors, because the inherent data is sparse so frame-based methods lead to unnecessarily high storage and computational requirements. In this work, we exhaustively compared with respect to accuracy and complexity different SNN models (SMLP, SLSTM and SGRU) on events captured in a IFC experiment. All models were properly designed in order to be implementation friendly for dedicated neuromorphic hardware. We demonstrate classification accuracy up to 98.43% for the SLSTM case, lower number of parameters equal to 606 for the SMLP case at 96% accuracy and a number of real multiplications varying from 5.035.600 for the SLSTM case to 61.200 for the Adaptive SMLP. In all benchmarks tests, the recurrent architectures prove to be more efficient in classification because they can benefit from the temporal relations across the timesteps. Taking into account that real multiplications can be substituted by a limited number of simple additions in the hardware level, the results of this work show that a potential co-integration of the event based camera with dedicated processor is really promising for real time and low complexity classification of events in flow cytometry.

**Disclosures.** The authors declare no conflict of interest.

**Data availability.** Data underlying the results presented in this paper are available in Ref. [29].